\documentclass[12pt]{article}
\usepackage{amsmath}
\usepackage{bm}
\usepackage{amssymb}
\usepackage{amsthm}
\usepackage{longtable}
\usepackage{amsfonts}
\usepackage{geometry}
\usepackage{multirow}
\usepackage{multicol}
\usepackage{graphicx}
\usepackage{epsfig}
\usepackage{epstopdf}
\usepackage{subfigure}
\usepackage{color}
\usepackage{ulem}
\usepackage{booktabs}
\usepackage{rotating}
\usepackage[toc,page]{appendix}
\usepackage{cite}
\usepackage{cases}
\usepackage{makecell}

\renewcommand{\theequation}{\arabic{section}.\arabic{equation}}
\renewcommand{\theequation}{\arabic{section}.\arabic{equation}}

\begin{document}
\def\a{\alpha}
\def\b{\beta}
\def\d{\delta}
\def\e{\epsilon}
\def\g{\gamma}
\def\h{\mathfrak{h}}
\def\k{\kappa}
\def\l{\lambda}
\def\o{\omega}
\def\p{\wp}
\def\r{\rho}
\def\t{\tau}
\def\s{\sigma}
\def\z{\zeta}
\def\x{\xi}
\def\V={{{\bf\rm{V}}}}
 \def\A{{\cal{A}}}
 \def\B{{\cal{B}}}
 \def\C{{\cal{C}}}
 \def\D{{\cal{D}}}
\def\K{{\cal{K}}}
\def\O{\Omega}
\def\R{\bar{R}}
\def\T{{\cal{T}}}
\def\L{\Lambda}
\def\f{E_{\tau,\eta}(sl_2)}
\def\E{E_{\tau,\eta}(sl_n)}
\def\Zb{\mathbb{Z}}
\def\Cb{\mathbb{C}}

\def\R{\overline{R}}

\def\beq{\begin{equation}}
\def\eeq{\end{equation}}
\def\bea{\begin{eqnarray}}
\def\eea{\end{eqnarray}}
\def\ba{\begin{array}}
\def\ea{\end{array}}
\def\no{\nonumber}
\def\le{\langle}
\def\re{\rangle}
\def\lt{\left}
\def\rt{\right}

\baselineskip=20pt

\renewcommand{\theequation}{\arabic{section}.\arabic{equation}}

\newcommand{\Title}[1]{{\baselineskip=26pt	\begin{center} \Large \bf #1  \\  \end{center}}}
\newcommand{\Author}{\begin{center}
	\large 
	Jiasheng Dong${}^{a}$,
    Pengcheng Lu${}^{b,c,}$\footnote{Corresponding author: lupengcheng@iphy.ac.cn},
    Junpeng Cao${}^{c,e,f,g}$,
    Wen-Li Yang${}^{a,g,h}$,
    \\ Ian Marquette${\,}^{d}$
    and Yao-Zhong Zhang${\,}^{b}$
\end{center}}

\newcommand{\Address}{\begin{center}
    ${}^a$ Institute of Modern Physics, Northwest University, Xian 710127, China\\
    ${}^b$ School of Mathematics and Physics, The University of Queensland, Brisbane, QLD 4072, Australia\\
	${}^c$ Beijing National Laboratory for Condensed Matter Physics, Institute of Physics, Chinese Academy of Sciences, Beijing 100190, China\\
    ${}^d$ Department of Mathematical and Physical Sciences, La Trobe University, Bendigo, VIC 3552, Australia\\
    ${}^e$ School of Physical Sciences, University of Chinese Academy of Sciences, Beijing 100049, China\\
	${}^f$ Songshan Lake Materials Laboratory, Dongguan, Guangdong 523808, China\\
	${}^g$ Peng Huanwu Center for Fundamental Theory, Xian 710127, China\\
	${}^h$ Shaanxi Key Laboratory for Theoretical Physics Frontiers, Xian 710127, China
\end{center}}

\Title{Root patterns and exact surface energy of the spin-1 Heisenberg model with generic open boundaries} 
\Author

\Address

\vspace{1truecm}

\begin{abstract}

We investigate the thermodynamic limit and exact surface energy of the isotropic spin-1 Heisenberg chain with integrable generic open boundary conditions by a novel Bethe ansatz method.
We obtain the homogeneous Bethe ansatz equations for the zero roots of the transfer matrix.
Based on the patterns of the zero roots, we analytical calculate the densities of zero roots and the surface energies of the model in all regimes of the boundary parameters.

\vspace{1truecm}

\noindent {\it Keywords}: Quantum spin chain; Bethe ansatz; Yang-Baxter equation
\end{abstract}
\newpage

\section{Introduction}
\label{intro} \setcounter{equation}{0}

The study of the quantum integrable models with open boundary conditions is an interesting and important subject because they describe systems with magnetic impurities or boundary magnetic fields \cite{JPA198720,JPA198821}.  In the recent years, open boundary integrable systems have also found many applications in open string theories \cite{JHEP201610,JPA201750,JHEP201905}, condensed matter physics \cite{SIGMA201713,JPA202053} and stochastic processes of non-equilibrium statistics \cite{Vanicat18,Godreau20,JPA202154}. For the quantum integrable models with diagonal boundary reflection matrices and $U(1)$ symmetry, the conventional coordinate and algebraic Bethe ansatz \cite{ZP1931205,CUP2014,SPD1978902,RMS197911,CUP1993} can be applied to solve the corresponding eigenvalue problems and a great number of papers have been devoted to this topic in the literature. However,  there exist large classes of open boundary integrable models without $U(1)$ symmetry \cite{NPB2003663,NPB1996478,NPB2002622,JPA200437} that can not be solved by the conventional Bethe ansatz methods.
In \cite{PRL2013137201,S2015}, the authors proposed a powerful off-diagonal Bethe ansatz method, which is applicable for solving integrable models with or without $U(1)$  symmetry. By means of this method, these authors and their collaborators have effectively solved many classes of integrable models without $U(1)$ symmetry.

Recently there has been a lot of interest in studying the thermodynamic limits of integrables models without $U(1)$ symmetry \cite{N}.  However, due to the inhomogeneonity of the Bethe ansatz equations (BAEs) and the complicated patterns of the Bethe roots for such models, the tradition thermodynamic Bethe ansatz (TBA) \cite{PRL19711301,TP1971401,CUP1005} could not be applied. This makes the study of the thermodynamic limits of these models very challenging.
The problem is solved by a novel $t-W$ method \cite{PRB2020085115,PRB2021L220401} which is effective for quantum integrable systems with or without $U(1)$ symmetry. The key point of this method is that the eigenvalue of the transfer matrix is parameterized by its zero points with well-defined patterns.

In this paper, we study the thermodynamic limit and surface energy of the isotropic spin-1 Heisenberg chain with non-diagonal boundary fields \cite{NPB2005711}. We obtain the homogeneous BAEs satisfied by the zero points of the eigenvalues of the fundamental and fused transfer matrices. The patterns of zero roots distributions in different regimes of the boundary parameters are determined by solving the BAEs and analytical analysis. Based on these results, we calculate the exact surface energies of all regimes in the thermodynamic limit. The method and process in this paper can be generalized to the study of the spin-$s$ Heisenberg chain model.

The paper is organized as follows.  Section 2 gives an introduction of the isotropic spin-1 Heisenberg chain with non-diagonal boundary fields and its integrability. The eigenvalues of the transfer matrices are parameterized by their zero points, and the homogeneous zero points BAEs are obtained. In section 3, we determine the patterns of zero roots distributions in different regimes. Based on the patterns, the thermodynamic limit and exact surface energies of all regimes are derived in section 4. In Section 5, we summarize our results and give some discussions. Some supporting materials are given in appendix A.


\section{Spin-1 Heisenberg model with non-diagonal boundary terms}
\label{spin1} \setcounter{equation}{0}
The spin-1 Heisenberg model with arbitrary boundary fields is described by the Hamiltonian
\bea
H = \frac{1}{\eta}\sum_{j=1}^{N-1} \left[\vec S_{j} \cdot \vec S_{j+1} -(\vec S_{j} \cdot \vec S_{j+1})^2 \right]+\frac{1}{\eta}\Big(3 N+\frac{8}{3}\Big)+H_{L}+H_{R} ,\label{H1}
\eea
where $N$ is the number of sites, $\eta$ is the crossing parameter, $\vec{S_{j}}(S^{x}_j,S^{y}_j,S^{z}_j)$ are the spin-1 operators on the $j$-th site, which have the matrix form in the 3-dimensional space $V$ of $sl(2)$,
\bea\label{Sj}
	S^{x}=\frac{1}{\sqrt{2}}\left(\begin{array}{ccc}
		0 & ~1 & ~0 \\
		1 & ~0 & ~1 \\
		0 & ~1 & ~0
	\end{array}\right),\hspace{0.15truecm}  S^{y}=\frac{1}{\sqrt{2}}\left(\begin{array}{ccc}
		0 & -i & 0 \\
		i & 0 & -i \\
		0 & i & 0
	\end{array}\right),\hspace{0.15truecm}  S^{z}=\left(\begin{array}{ccc}
		1 & ~0 & ~0 \\
		0 & ~0 & ~0 \\
		0 & ~0 & -1
	\end{array}\right),\\\nonumber
\eea
and $H_{L}$, $H_{R}$ denote  the left and right boundary fields, respectively,
\bea
H_{L}=&&\hspace{-0.5truecm}\frac{1}{p_{-}^2-\frac{1}{4}\left(1+\alpha_{-}^2\right) \eta^2}\bigg[2 p_{-}\left(\alpha_{-} \cos \phi_{-} S_1^x-\alpha_{-} \sin \phi_{-} S_1^y+S_1^z\right)-\eta\left(S_1^z\right)^2 \no\\[6pt]
&&\hspace{-0.5truecm} -\frac{1}{2} \alpha_{-}^2 \eta\left[\cos \left(2 \phi_{-}\right)\left[\left(S_1^x\right)^2-\left(S_1^y\right)^2\right]-\left(S_1^z\right)^2\right]-\alpha_{-} \eta \cos \phi_{-}\left[S_1^x S_1^z+S_1^z S_1^x\right] \no\\[6pt]
&&\hspace{-0.5truecm} +\frac{1}{2} \alpha_{-}^2 \eta \sin \left(2 \phi_{-}\right)\left[S_1^x S_1^y+S_1^y S_1^x\right]+\alpha_{-} \eta \sin \phi_{-}\left[S_1^y S_1^z+S_1^z S_1^y\right]+\eta I_1\bigg],\\[6pt]
H_{R}=&&\hspace{-0.5truecm}\frac{1}{p_{+}^2-\frac{1}{4}\left(1+\alpha_{+}^2\right) \eta^2}\bigg[2 p_{+}\left(\alpha_{+} \cos \phi_{+} S_N^x-\alpha_{+} \sin \phi_{+} S_N^y-S_N^z\right)-\eta\left(S_N^z\right)^2 \no\\[6pt]
&&\hspace{-0.5truecm} -\frac{1}{2} \alpha_{+}^2 \eta\left[\cos \left(2 \phi_{+}\right)\left[\left(S_N^x\right)^2-\left(S_N^y\right)^2\right]-\left(S_N^z\right)^2\right]+\alpha_{+} \eta \cos \phi_{+}\left[S_N^x S_N^z+S_N^z S_N^x\right]\no \\[6pt]
&&\hspace{-0.5truecm} +\frac{1}{2} \alpha_{+}^2 \eta \sin \left(2 \phi_{+}\right)\left[S_N^x S_N^y+S_N^y S_N^x\right]-\alpha_{+} \eta \sin \phi_{+}\left[S_N^y S_N^z+S_N^z S_N^y\right]+\eta I_N\bigg].
\eea
In the above expressions, $p_{\pm}$, $\alpha_{\pm}$ and $\phi_{\pm}$ are boundary parameters which measure the strength and direction of the boundary fields. The hermiticity of the Hamiltonian (\ref{H1}) requires that the crossing parameter and the boundary parameters are real. Note that the boundary fields are unparallel and thus the $U(1)$ symmetry of the system is broken.

\subsection{Integrability of the model}
\label{Integrability}
The Hamiltonian (\ref{H1}) is constructed by using the spin-$(1,1)$ $R$-matrix $R^{\left(1, 1\right)}$ and the reflection matrices $K^{\pm(1)}$ based on the quantum inverse scattering method. The $R$-matrix $R^{\left(1, 1\right)}$ defined in the tensor space $V\otimes V$ is
\bea\label{R11}
R_{12}^{\left(1, 1\right)}(u)=\left(\begin{array}{ccc|ccc|ccc}
c(u) & & & & & & & & \\
& b(u) & & e(u) & & & & & \\
& & d(u) & & g(u) & & f(u) & & \\ \hline
& e(u) & & b(u) & & & & & \\
& & g(u) & & a(u) & & g(u) & & \\
& & & & & b(u) & & e(u) & \\\hline
& & f(u)&  &g(u)&  &d(u) &  & \\
& & & & & e(u) & & b(u) & \\
& & & & & & & & c(u)
\end{array}\right),
\eea
with the non-zero entries
\bea
& a(u)=u(u+\eta)+2 \eta^2, \quad b(u)=u(u+\eta), \quad c(u)=(u+\eta)(u+2 \eta), \no \\
& d(u)=u(u-\eta), \quad e(u)=2 \eta(u+\eta), \quad f(u)=2 \eta^2, \quad g(u)=2 u \eta .
\eea
The $R$-matrix satisfies the quantum Yang-Baxter equation (QYBE) \cite{PRL196719,Baxter1982}
\bea\label{QYBE}
R_{12}^{\left(1, 1\right)}(u-v) R_{13}^{\left(1, 1\right)}(u) R_{23}^{\left(1, 1\right)}(v)=R_{23}^{\left(1, 1\right)}(v) R_{13}^{\left(1, 1\right)}(u) R_{12}^{\left(1, 1\right)}(u-v).
\eea
The generic non-diagonal $K$-matrix $K^{-(1)}(u))$ in the space $V$ is given by \cite{PLA1990147,NPB1996470}
\bea\label{K11f}
K^{-(1)}(u)=(2 u+\eta)\left(\begin{array}{lll}
x_1(u) & y_4^{\prime}(u) & y_6^{\prime}(u) \\
y_4(u) & x_2(u) & y_5^{\prime}(u) \\
y_6(u) & y_5(u) & x_3(u)
\end{array}\right)
\eea
with
\bea
&& x_1(u)=\left(p_{-}+u+\frac{\eta}{2}\right)\left(p_{-}+u-\frac{\eta}{2}\right)+\frac{\alpha_{-}^2}{2} \eta\left(u-\frac{\eta}{2}\right), \no\\[6pt]
&& x_2(u)=\left(p_{-}+u-\frac{\eta}{2}\right)\left(p_{-}-u+\frac{\eta}{2}\right)+\alpha_{-}^2\left(u+\frac{\eta}{2}\right)\left(u-\frac{\eta}{2}\right), \no\\[6pt]
&& x_3(u)=\left(p_{-}-u-\frac{\eta}{2}\right)\left(p_{-}-u+\frac{\eta}{2}\right)+\frac{\alpha_{-}^2}{2} \eta\left(u-\frac{\eta}{2}\right), \no\\[6pt]
&& y_4(u)=\sqrt{2} \alpha_{-} e^{-i \phi_{-}} u\left(p_{-}+u-\frac{\eta}{2}\right), \quad y_4^{\prime}(u)=\sqrt{2} \alpha_{-} e^{i \phi_{-}} u\left(p_{-}+u-\frac{\eta}{2}\right), \no\\[6pt]
&& y_5(u)=\sqrt{2} \alpha_{-} e^{-i \phi_{-}} u\left(p_{-}-u+\frac{\eta}{2}\right), \quad y_5^{\prime}(u)=\sqrt{2} \alpha_{-} e^{i \phi_{-}} u\left(p_{-}-u+\frac{\eta}{2}\right), \no\\[6pt]
&& y_6(u)=\alpha_{-}^2 e^{-2 i \phi_{-}} u\left(u-\frac{\eta}{2}\right), \quad y_6^{\prime}(u)=\alpha_{-}^2 e^{2 i \phi_{-}} u\left(u-\frac{\eta}{2}\right),
\eea
which is the generic solution of the corresponding reflection equation (RE)
\bea\label{RE}
\hspace{-0.1truecm}R_{12}^{\left(1, 1\right)}(u\hspace{-0.07truecm}-\hspace{-0.07truecm}v) K_1^{-(1)}(u) R_{21}^{\left(1, 1\right)}(u\hspace{-0.07truecm}+\hspace{-0.07truecm}v) K_2^{-(1)}(v)\hspace{-0.07truecm}=\hspace{-0.07truecm}K_2^{-(1)}(v) R_{21}^{\left(1, 1\right)}(u\hspace{-0.07truecm}+\hspace{-0.07truecm}v) K_1^{-(1)}(u) R_{12}^{\left(1, 1\right)}(u\hspace{-0.07truecm}-\hspace{-0.07truecm}v),
\eea
where $K_1^{-(1)}=K^{-(1)}\otimes 1$, $K_2^{-(1)}=1\otimes K^{-(1)}$, $R_{21}^{\left(1, 1\right)}=P_{12}R_{12}^{\left(1, 1\right)}P_{12}$ and  $P_{12}= R_{12}^{\left(1, 1\right)}(0)/(2\eta^2)$ is the permutation operator.
The dual reflection matrix $K^{+(1)}(u)$ is given by the duality
\bea\label{K11z}
K^{+(1)}(u)=\left.K^{-(1)}(-u-\eta)\right|_{\left(p_{-}, \alpha_{-}, \phi_{-}\right) \rightarrow\left(p_{+}, \alpha_{+}, \phi_{+}\right)},
\eea
satisfying the dual RE
\bea\label{DUAL-RE}
&& R_{12}^{\left(1, 1\right)}(v-u) K_1^{+(1)}(u) R_{21}^{\left(1, 1\right)}(-u-v-2 \eta) K_2^{+(1)}(v) \no\\[6pt]
&& \qquad\qquad = K_2^{+(1)}(v) R_{21}^{\left(1, 1\right)}(-u-v-2 \eta) K_1^{+(1)}(u) R_{12}^{\left(1, 1\right)}(v-u).
\eea
The spin-$(1,1)$ single row monodromy matrix $T_0^{\left(1, 1\right)}(u)$ and the reflecting monodromy matrix  $\hat{T}_0^{\left(1, 1\right)}(u)$ of the system are constructed by the $R^{\left(1, 1\right)}$-matrix as
\bea
&& T_0^{\left(1, 1\right)}(u)=R_{0 N}^{\left(1, 1\right)}\left(u-\theta_N\right) R_{0 N-1}^{\left(1, 1\right)}\left(u-\theta_{N-1}\right) \cdots R_{01}^{\left(1, 1\right)}\left(u-\theta_1\right), \no\\[6pt]
&& \hat{T}_0^{\left(1, 1\right)}(u)=R_{10}^{\left(1, 1\right)}\left(u+\theta_1\right) R_{20}^{\left(1, 1\right)}\left(u+\theta_2\right) \cdots R_{N 0}^{\left(1, 1\right)}\left(u+\theta_N\right),\no
\eea
where $\{\theta_j |j=1, \ldots, N\}$ are the inhomogeneous parameters,  $T_0^{\left(1, 1\right)}(u)$ and $\hat{T}_0^{\left(1, 1\right)}(u)$ are the $3\times3$ matrices in the auxiliary space $V_0$ and their elements act on the $N$-fold quantum space $V\otimes V\otimes\cdots\otimes V$. The spin-$(1,1)$ transfer matrix  is
\bea\label{t11}
t^{\left(1, 1\right)}(u)=t r_0\left\{K_0^{+(1)}(u) T_0^{\left(1, 1\right)}(u) K_0^{-(1)}(u) \hat{T}_0^{\left(1, 1\right)}(u)\right\},
\eea
where $tr_0$ means the partial trace over the auxiliary space. The Hamiltonian (\ref{H1}) is generated from the transfer matrix as
\bea
H= \left.\frac{\partial\ln t^{\left(1, 1\right)}(u)}{\partial u}\right|_{u=0,\left\{\theta_j=0\right\}}.
\eea
The QYBE (\ref{QYBE}),the RE (\ref{RE}) and its dual (\ref{DUAL-RE}) ensure the integrability of the model (\ref{H1}).

\subsection{Exact solution}
\label{Exact}
To solve the Hamiltonian (\ref{H1}), we also need to introduce the fundamental spin-$(\frac{1}{2},1)$ transfer matrix $t^{\left(\frac{1}{2}, 1\right)}(u)$ defined as
\bea\label{t1-21}
t^{\left(\frac{1}{2}, 1\right)}(u)=t r_{\bar {0}}\left\{K_{\bar {0}}^{+(\frac{1}{2})}(u) T_{\bar {0}}^{\left(\frac{1}{2}, 1\right)}(u) K_{\bar {0}}^{-(\frac{1}{2})}(u) \hat{T}_{\bar {0}}^{\left(\frac{1}{2}, 1\right)}(u)\right\},
\eea
where we have used the subscript $\bar{0}$ to denote the two-dimensional auxiliary space. In appendix A, we have provided some useful properties of the transfer matrix $t^{\left(\frac{1}{2}, 1\right)}(u)$ (\ref{t1-21}) and the corresponding spin-$(\frac{1}{2},1)$ $R$-matrix $R^{\left(\frac{1}{2}, 1\right)}(u)$.

 Using the properties of the $R$-matrices $R^{\left(1, 1\right)}(u)$ and $R^{\left(\frac{1}{2}, 1\right)}(u)$, the following relations can be easily proved
 \bea
 &&\hspace{-1.5truecm}t^{\left(1, 1\right)}(u)=t^{\left(1, 1\right)}(-u-\eta),\quad t^{\left(\frac{1}{2}, 1\right)}(u)=t^{\left(\frac{1}{2}, 1\right)}(-u-\eta),\\[6pt]
 &&\hspace{-1.5truecm}t^{(\frac{1}{2},1)}(0)=2p_{-}p_{+}\prod_{l=1}^{N}(\theta_l+\frac{3}{2}\eta)(-\theta_l+\frac{3}{2}\eta)\times{\rm id},\\[6pt]
 &&\hspace{-1.5truecm}t^{(\frac{1}{2},1)}(u)|_{u\rightarrow\infty}=2(\alpha_{-}\alpha_{+}-1)u^{2N+2}\times{\rm id}+\cdots,\\[6pt]
 &&\hspace{-1.5truecm}t^{1,1)}(u)|_{u\rightarrow\infty}=4[(1+\alpha_{+}^2)(1+\alpha_{-}^2)-4(\alpha_{+}\alpha_{-}-1)^2] u^{4N+6}\times{\rm id}+\cdots,\\[6pt]
 &&\hspace{-1.5truecm} t^{\left(1, 1\right)}(u)  =-4 u(u+\eta) t^{\left(\frac{1}{2}, 1\right)}\left(u+\frac{\eta}{2}\right) t^{\left(\frac{1}{2}, 1\right)}\left(u-\frac{\eta}{2}\right)+4 u(u+\eta) \delta^{(1)}\left(u+\frac{\eta}{2}\right),\\[6pt]
&&\hspace{-1.5truecm}t^{\left(1, 1\right)}\left(\theta_j\right) t^{\left(\frac{1}{2}, 1\right)}(\theta_j-\frac{3 \eta}{2})=-4 \theta_j\left(\theta_j+\eta\right) \delta^{(1)}(\theta_j-\frac{\eta}{2}) t^{\left(\frac{1}{2}, 1\right)}(\theta_j+\frac{\eta}{2}),\quad j=1, \ldots, N,
 \eea
where the coefficient function $\delta^{(1)}(u)$ (which is related to the quantum determinant) is
\bea\label{ad}
\delta^{(1)}(u) \hspace{-0.2truecm}&=&\hspace{-0.2truecm}a^{(1)}(u) d^{(1)}(u-\eta), \no\\[6pt]
a^{(1)}(u)\hspace{-0.2truecm}& =&\hspace{-0.2truecm}d^{(1)}(-u-\eta) \no\\[6pt]
\hspace{-0.2truecm}& =&\hspace{-0.2truecm}-\frac{2 u+2 \eta}{2 u+\eta}\left(\sqrt{1+\alpha_{+}^2} u+p_{+}\right)\left(\sqrt{1+\alpha_{-}^2} u-p_{-}\right) \prod_{l=1}^N\left(u \pm \theta_l+\frac{3 \eta}{2}\right).
\eea
From the definitions (\ref{t11}) and (\ref{t1-21}), we know that the transfer matrices $t^{\left(1, 1\right)}(u)$ and $t^{\left(\frac{1}{2}, 1\right)}(u)$, as the functions of $u$, are operator-valued polynomials of degree $4N+6$ and $2N+2$ respectively. Let $|\Psi\rangle$ denote the common eigenstate of the transfer matrices with eigenvalues $\Lambda^{(1,1)}(u)$ and $\Lambda^{(\frac{1}{2},1)}(u)$, respectively. Namely,
\bea
t^{(1,1)}(u)|\Psi\rangle=\Lambda^{(1,1)}\Psi\rangle, \qquad t^{(\frac{1}{2},1)}(u)|\Psi\rangle=\Lambda^{(\frac{1}{2},1)}\Psi\rangle.
\eea
From the above analysis, we know that the eigenvalues $\Lambda^{(1,1)}(u)$ and $\Lambda^{(\frac{1}{2},1)}(u)$ satisfy
 \bea
 &&\hspace{-1.5truecm}\Lambda^{\left(1, 1\right)}(u)=t^{\left(1, 1\right)}(-u-\eta),\quad t^{\left(\frac{1}{2}, \label{cross} 1\right)}(u)=\Lambda^{\left(\frac{1}{2}, 1\right)}(-u-\eta),\\[6pt]
 &&\hspace{-1.5truecm}\Lambda^{(\frac{1}{2},1)}(0)=2p_{-}p_{+}\prod_{l=1}^{N}(\theta_l+\frac{3}{2}\eta)(-\theta_l+\frac{3}{2}\eta),\\[6pt]
 &&\hspace{-1.5truecm}\Lambda^{(\frac{1}{2},1)}(u)|_{u\rightarrow\infty}=2(\alpha_{-}\alpha_{+}-1)u^{2N+2}+\cdots,\\[6pt]
 &&\hspace{-1.5truecm}\Lambda^{(1,1)}(u)|_{u\rightarrow\infty}=4[(1+\alpha_{+}^2)(1+\alpha_{-}^2)-4(\alpha_{+}\alpha_{-}-1)^2] u^{4N+6}+\cdots,\\[6pt]
 &&\hspace{-1.5truecm} \Lambda^{\left(1, 1\right)}(u)  =-4 u(u+\eta) \Lambda^{\left(\frac{1}{2}, 1\right)}\left(u+\frac{\eta}{2}\right) \Lambda^{\left(\frac{1}{2}, 1\right)}\left(u-\frac{\eta}{2}\right)+4 u(u+\eta) \delta^{(1)}\left(u+\frac{\eta}{2}\right),\label{lam11}\\[6pt]
&&\hspace{-1.5truecm}\Lambda^{\left(1, 1\right)}\left(\theta_j\right) \Lambda^{\left(\frac{1}{2}, 1\right)}(\theta_j-\frac{3 \eta}{2})=-4 \theta_j\left(\theta_j+\eta\right) \delta^{(1)}(\theta_j-\frac{\eta}{2}) \Lambda^{\left(\frac{1}{2}, 1\right)}(\theta_j+\frac{\eta}{2}),\quad j=1, \ldots, N,\label{hds}
 \eea
The above relations allow us to express the transfer matrix $\Lambda^{(1,1)}(u)$ (resp.  the fundamental one $\Lambda^{(\frac{1}{2},1)}(u)$) in terms of its $4N+6$ zero points $\{\pm z^{(1)}_k|k=1,\ldots,2N+3\}$ (resp.  $2N+2$ zero points $\{\pm z_l|l=1,\ldots,N+1\}$) as follows
\bea\label{lamz}
\Lambda^{\left(1, 1\right)}(u) \hspace{-0.6truecm}&& =\Lambda_0 \prod_{k=1}^{2 N+3}\left(u-z_k^{(1)}+\frac{\eta}{2}\right)\left(u+z_k^{(1)}+\frac{\eta}{2}\right) ,\no\\[6pt]
\Lambda^{\left(\frac{1}{2}, 1\right)}(u)\hspace{-0.6truecm} && =2(\alpha_{-}\alpha_{+}-1) \prod_{l=1}^{N+1}\left(u-z_l+\frac{\eta}{2}\right)\left(u+z_l+\frac{\eta}{2}\right),
\eea
where the coefficient $\Lambda_0=4[(1+\alpha_{+}^2)(1+\alpha_{-}^2)-4(\alpha_{+}\alpha_{-}-1)^2]$. Substituting the parameterization (\ref{lamz}) into (\ref{lam11})-(\ref{hds}),  we can obtain the homogeneous zero points BAEs. Thus the relations (\ref{cross})-(\ref{hds}) allow us to completely determine the $3N+5$ unknowns $\{z^{(1)}_k\}$ and $\{z_l\}$. In terms of zero roots $\{z^{(1)}_k\}$, the energy spectrum of the Hamiltonian (\ref{H1}) is expressed as
\bea
E=-\sum_{k=1}^{2N+3}\frac{\eta}{(z_k^{(1)})^2-\frac{\eta^2}{4}}.
\eea

\section{Patterns of zero roots}
\label{patterns} \setcounter{equation}{0}
Without loss of generality, we choose all inhomogeneity parameters to be imaginary, $\{\theta_j\equiv i\bar{\theta}_j\}$, and let $\{\bar{z}^{(1)}_k\equiv -iz^{(1)}_k,\,\bar{z}_l\equiv -iz_l\}$.
Putting $u=i\bar{z}_l$ and $u=-i\bar{z}_l$ into (\ref{lam11}) and dividing one resulting equation by another one, in the homogeneous limit $\{\theta_j\rightarrow 0\}$, we have
 \bea\label{z1zBA}
 \Big[\frac{(\bar{z}_l-2i)(\bar{z}_l+i)}{(\bar{z}_l+2i )(\bar{z}_l-i)}  \Big]^{2N}=&&\hspace{-0.4truecm}\frac{\bar{z}_l+\frac{3i}{2}}{\bar{z}_l-\frac{3i}{2}}\frac{\bar{z}_l-\frac{i}{2}}{\bar{z}_l+\frac{i}{2}}\frac{\bar{z}_l-ip}{\bar{z}_l+ip}\frac{\bar{z}_l+ip+i}{\bar{z}_l-ip-i}\frac{\bar{z}_l-iq}{\bar{z}_l+iq}\frac{\bar{z}_l+iq+i}{\bar{z}_l-iq-i}\no\\[6pt]
 &&\hspace{-0.4truecm}\times\prod_{k=1}^{2N+3}\frac{\bar{z}_l-\bar{z}^{(1)}_k-\frac{i}{2}}{\bar{z}_l-\bar{z}^{(1)}_k+\frac{i}{2}}\frac{\bar{z}_l+\bar{z}^{(1)}_k-\frac{i}{2}}{\bar{z}_l+\bar{z}^{(1)}_k+\frac{i}{2}},
 \eea
where we have set $\eta=1$, $p=\frac{p_+}{\sqrt{1+\alpha^2_{+}}}-\frac{1}{2}$, and $q=-\frac{p_-}{\sqrt{1+\alpha^2_{-}}}-\frac{1}{2}$ for convenience.
For a complex zero root $\bar{z}_l$ with a negative imaginary part, we readily have
\bea
\Big|(\bar{z}_l-2i)(\bar{z}_l+i)\Big|>\Big|(\bar{z}_l+2i)(\bar{z}_l-i)\Big|.
\eea
This indicates that the module of the left hand side of Eq.(\ref{z1zBA}) is larger than 1. Thus in the thermodynamic limit $N\rightarrow \infty$, the left hand side tends to infinity exponentially. To keep Eq.(\ref{z1zBA}) holding, the right hand side of Eq.(\ref{z1zBA}) must also tend to infinity in the same order.  Thus the denominator of the first term in the right hand side must tend to zero exponentially, which leads to
\bea\label{z1zrelation}
\bar{z}_l-\bar{z}^{(1)}_k+\frac{i}{2}\rightarrow 0\quad {\rm and}\quad \bar{z}_l+\bar{z}^{(1)}_k+\frac{i}{2}\rightarrow 0.
\eea
These results also appear in the exact numerical diagonalization results below and will help us to study the surface energy.

\begin{figure}[htbp]
\centering
\includegraphics[width=15cm]{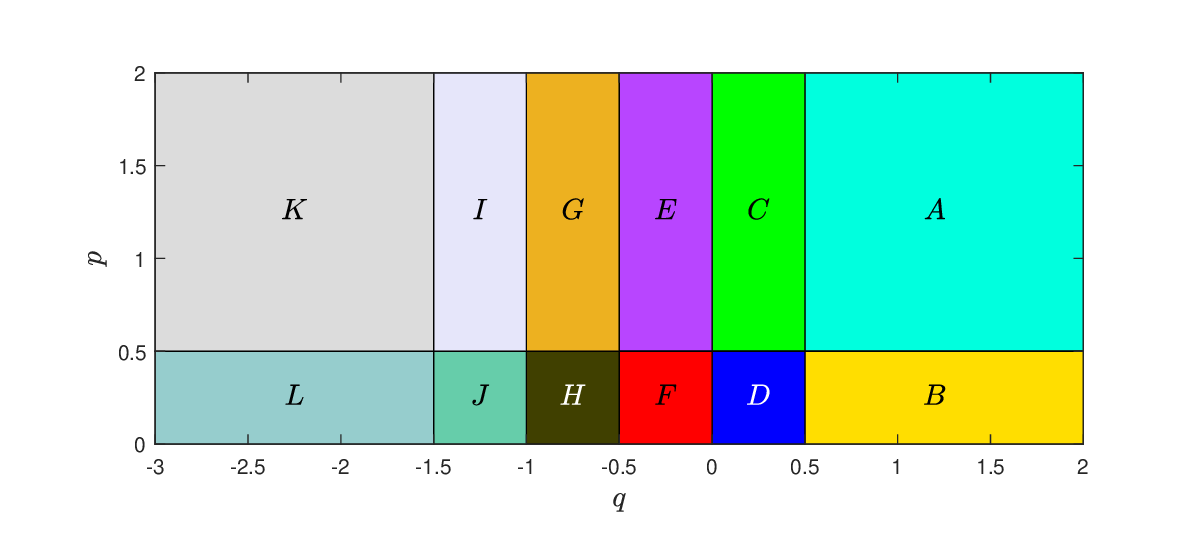}
  \caption{The distribution of $\bar{z}^{(1)}$-roots and $\bar{z}$-roots at the ground state in the upper $p-q$ plane for fixed  $\eta=1$.}
\label{diagram-spin1}
\end{figure}

We now study the solutions of zero roots $\{\bar{z}^{(1)}_k\}$ and $\{\bar{z}_l\}$ at the ground state. Through numerical calculation and algebraic analysis, we find that the distribution of the zero roots for the ground state can be divided into 12 different regimes in the upper $p-q$ plane, as shown in Fig.\ref{diagram-spin1}. We should note that
the results in the lower $p-q$ plane can be obtained directly from the symmetry of the Hamiltonian, without further calculation.

\begin{figure}[htbp]
\center
\includegraphics[width=12cm]{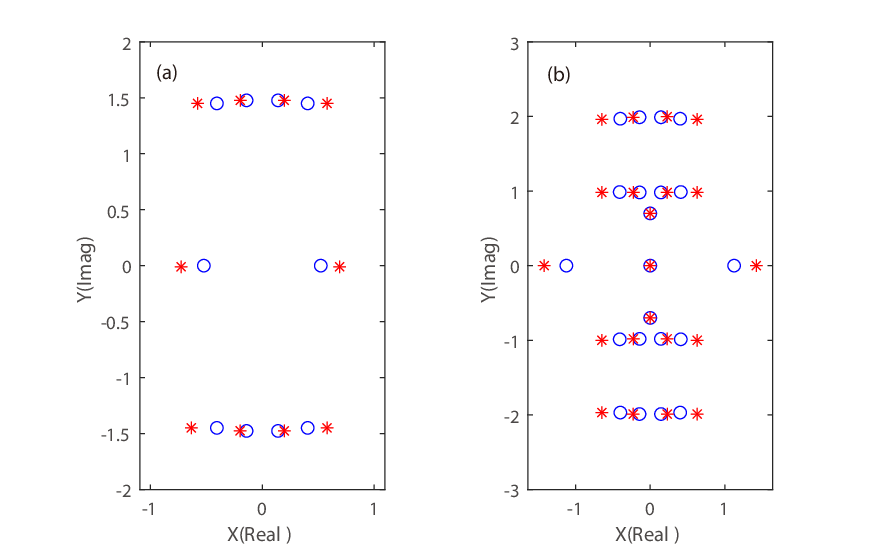}
  \caption{Exact numerical diagonalization results of the zero roots distributions at the ground state in region E with $N=4,p=0.6,q=-0.2$. (a) the $\bar{z}$-roots of the eigenvalue $\Lambda^{\left(\frac{1}{2}, 1\right)}(u)$; (b) the $\bar{z}^{(1)}$-roots of the eigenvalue $\Lambda^{\left(1, 1\right)}(u)$. The blue
   circles indicate the roots for $\{\bar{\theta}_j=0|j=1,\ldots,2N\}$ and the red asterisks specify the roots
   with the inhomogeneity parameters $\{\bar{\theta}_j=0.1 (j-N-0.5)|j=1,\ldots,2N\}$.}
\label{ground1-ospin1}
\end{figure}

\begin{figure}[htbp]
\center
\includegraphics[width=12cm]{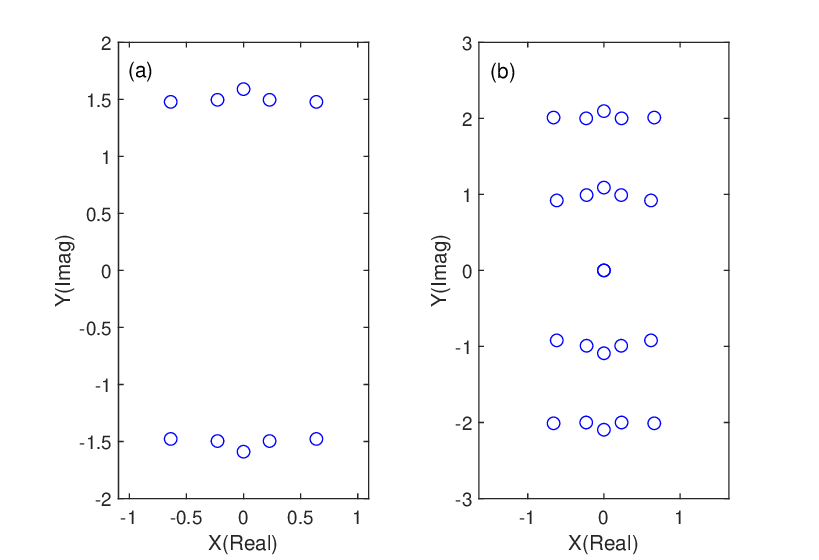}
  \caption{Exact numerical diagonalization results of the zero roots distributions at the ground state in region K with $N=4,p=0.6,q=-2.5$. (a) the $\bar{z}$-roots of the eigenvalue $\Lambda^{\left(\frac{1}{2}, 1\right)}(u)$; (b) the $\bar{z}^{(1)}$-roots of the eigenvalue $\Lambda^{\left(1, 1\right)}(u)$.}
\label{ground1-ospin2}
\end{figure}

\begin{figure}[htbp]
\center
\includegraphics[width=12cm]{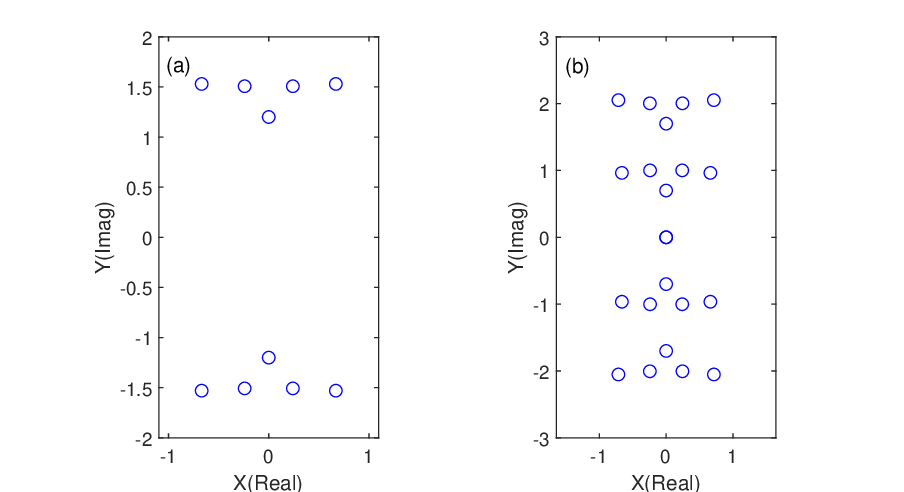}
  \caption{Exact numerical diagonalization results of the zero roots distributions at the ground state in region I with $N=4,p=1.5,q=-1.2$. (a) the $\bar{z}$-roots of the eigenvalue $\Lambda^{\left(\frac{1}{2}, 1\right)}(u)$; (b) the $\bar{z}^{(1)}$-roots of the eigenvalue $\Lambda^{\left(1, 1\right)}(u)$.}
\label{ground1-ospin3}
\end{figure}

\noindent The patterns of the zero roots distributions in 12 different regimes are listed  in Tab.\ref{open-spin1-z-table}. In addition, we also show some exact numerical diagonalization results in Figs.\ref{ground1-ospin1}-\ref{ground1-ospin3}.
\begin{table}\renewcommand\arraystretch{2.1}

\resizebox{\textwidth}{!}{
\begin{tabular}{|c|c|c|c|}
\hline & $\bar{z}$-roots & $\bar{z}^{(1)}$-roots & $j_{max},k_{max}$\\

\hline $\mathrm{A}$ & $z_0, \tilde{z}_j+\frac{3}{2} i$ & $0, z_1, z_2, \tilde{z}^{(1)}_k+i, \tilde{z}^{(1)}_k+2 i$ & $N$\\

\hline $\mathrm{B}$ & $z_0, z_x i,(1+p) i, \tilde{z}_j+\frac{3}{2} i$. & \makecell{$0, z_1, z_2,\left(z_x-\frac{1}{2}\right) i,\left(z_x+\frac{1}{2}\right) i,\left(\frac{1}{2}+p\right) i,\left(\frac{3}{2}+p\right) i$,\\ $\tilde{z}^{(1)}_k+i, \tilde{z}^{(1)}_k+2 i$} & $ N-2$\\
\hline $\mathrm{C}$ & $z_0, z_x i,(1+q) i, \tilde{z}_j+\frac{3}{2} i$. & \makecell{$0, z_1, z_2,\left(z_x-\frac{1}{2}\right) i,\left(z_x+\frac{1}{2}\right) i,\left(\frac{1}{2}+q\right) i,\left(\frac{3}{2}+q\right) i$,\\$ \tilde{z}^{(1)}_k+i, \tilde{z}^{(1)}_k+2 i$} & $N-2$\\
\hline $\mathrm{D}$ & $z_0,(1+p) i,(1+q) i, \tilde{z}_j+\frac{3}{2} i$. &\makecell{ $0, z_1, z_2,\left(\frac{1}{2}+p\right) i,\left(\frac{3}{2}+p\right) i,\left(\frac{1}{2}+q\right) i,\left(\frac{3}{2}+q\right) i$,\\$ \tilde{z}^{(1)}_k+i, \tilde{z}^{(1)}_k+2 i$} & $N-2$\\
\hline $\mathrm{E}$ & $z_0, \tilde{z}_j+\frac{3}{2} i$ & $0, z_1,\left(\frac{1}{2}+|q|\right) i, \tilde{z}^{(1)}_k+i, \tilde{z}^{(1)}_k+2 i$ & $N$\\
\hline $\mathrm{F}$ & $z_0, z_x i,(1+p) i, \tilde{z}_j+\frac{3}{2} i$ &\makecell{ $0, z_1,\left(z_x-\frac{1}{2}\right) i,\left(z_x+\frac{1}{2}\right) i,\left(\frac{1}{2}+p\right) i,\left(\frac{3}{2}+p\right) i,\left(\frac{1}{2}+|q|\right) i$,\\$ \tilde{z}^{(1)}_k+i, \tilde{z}^{(1)}_k+2 i$} & $N-2$\\
\hline $\mathrm{G}$ & $z_x i, \lambda+i,-\lambda+i, \tilde{z}_j+\frac{3}{2} i$ & \makecell{$0, z_1,\left(z_x-\frac{1}{2}\right) i,\left(z_x+\frac{1}{2}\right) i, \lambda+\frac{3}{2} i,-\lambda+\frac{3}{2} i,\left(\frac{3}{2}-|q|\right) i$,\\$ \tilde{z}^{(1)}_k+i, \tilde{z}^{(1)}_k+2 i$}& $N-2$ \\
\hline $\mathrm{H}$ & $(1+p) i, \lambda+i,-\lambda+i, \tilde{z}_j+\frac{3}{2} i$ &\makecell{ $0, z_1,\left(\frac{1}{2}+p\right) i,\left(\frac{3}{2}+p\right) i,\left(\frac{3}{2}-|q|\right) i, \lambda+\frac{3}{2} i,-\lambda+\frac{3}{2} i$,\\$ \tilde{z}^{(1)}_k+i, \tilde{z}^{(1)}_k+2 i$}& $N-2$ \\
\hline $\mathrm{I}$ & $|q| i, \tilde{z}_j+\frac{3}{2} i$ & $0,\left(|q|-\frac{1}{2}\right) i,\left(|q|+\frac{1}{2}\right) i, \tilde{z}^{(1)}_k+i, \tilde{z}^{(1)}_k+2 i$ & $N$\\

\hline $\mathrm{J}$ & $z_x i,|q| i,(1+p) i, \tilde{z}_j+\frac{3}{2} i$ & \makecell{ $0,\left(z_x-\frac{1}{2}\right) i,\left(z_x+\frac{1}{2}\right) i,\left(|q|-\frac{1}{2}\right) i,\left(|q|+\frac{1}{2}\right) i$,\\$\left(\frac{1}{2}+p\right) i,\left(\frac{3}{2}+p\right) i, \tilde{z}^{(1)}_k+i, \tilde{z}^{(1)}_k+2 i$ }& $N-2$\\

\hline $\mathrm{K}$ & $z_x i, \tilde{z}_j+\frac{3}{2} i$ & $0,\left(z_x-\frac{1}{2}\right) i,\left(z_x+\frac{1}{2}\right) i, \tilde{z}^{(1)}_k+i, \tilde{z}^{(1)}_k+2 i$ & $ N$\\
\hline $\mathrm{L}$ & $(1+p) i, \tilde{z}_j+\frac{3}{2} i$. & $0,\left(\frac{1}{2}+p\right) i,\left(\frac{3}{2}+p\right) i, \tilde{z}^{(1)}_k+i, \tilde{z}^{(1)}_k+2 i$ & $ N$\\
\hline
\end{tabular}
}
\caption{Patterns of $\bar{z}$-roots and $\bar{z}^{(1)}$-roots distribution at the ground state in 12 different regimes.
Here, $z_x$ is a real number larger than $\frac{3\eta}{2}$ and $z_0, z_1, z_2$ would tend to infinity in the thermodynamic limit.}
\label{open-spin1-z-table}
\end{table}

\newpage
As shown in Fig.\ref{ground1-ospin1}, the choice of the pure real $\{\bar{\theta}_j\}$ does not change the patterns of the roots $\{\bar{z}_j\}$ but the roots density, which allows us to calculate the physical quantities such as the surface energy of the system in the thermodynamic limit with the help of suitable $\{\bar{\theta}_j\}$ \cite{PRB2021L220401}.

\section{Surface energy}\label{Interaction-case1}
\setcounter{equation}{0}
Based on the patterns of the zero roots distribution in Tab.\ref{open-spin1-z-table}, we can now study the surface energy induced by the boundaries. The surface energy is defined by $E_b=E_g-E_p$, where $E_g$ is the ground state energy of the system and $E_p=-N$ is the ground state energy of the corresponding periodic chain.

Without losing generality, we choose the regime $B$ as an example to show the process, where $ 0<p<\frac{1}{2}$ and $q>\frac{1}{2}$. The $\bar{z}$-roots of the eigenvalue $\Lambda^{\left(\frac{1}{2}, 1\right)}(u)$ form a set of two strings $\{\bar{z}_l\sim \tilde{z}_l\pm \frac{3i}{2}|l=1,\ldots,N\}$, one pair of boundary string $\pm(1+p)i$ and some extra roots;  the $\bar{z}^{(1)}$-roots of the eigenvalue $\Lambda^{\left(1, 1\right)}(u)$ form a set of four strings $\{\bar{z}_k\sim \tilde{z}_k\pm ni|k=1,\ldots,N;n=1,2\}$, two pairs of boundary strings $\pm(1+p)i, \pm(\frac{3}{2}+p)i$ and some extra roots in Tab.\ref{open-spin1-z-table}.
In the thermodynamic limit, the distributions of $\{\tilde{z}_l\}$ and $\{\tilde{z}_k^{(1)}\}$ can be characterized by the densities $\rho(\tilde{z})$ and $\rho(\tilde{z}^{(1)})$, respectively. From the constraint (\ref{z1zrelation}), we obtain $\rho(\tilde{z})=\rho(\tilde{z}^{(1)})$.

Furthermore, we assume that the density of inhomogeneity parameters $1/[N(\bar{\theta}_j-\bar{\theta}_{j-1} )]$ has the continuum limit $\sigma(\bar{\theta})$.
Taking the logarithm of the Eq.(\ref{hds}) and subtracting $\bar{\theta}_{j-1}$ from $\bar{\theta}_j$, by omitting the $O(N^{-1})$ terms we readily have
\bea\label{intlam}
&&\hspace{-0.4truecm} N \int_{-\infty}^{\infty}\left[2 b_{3}(u-\tilde{z})+b_{1}(u-\tilde{z})+b_{5}(u-\tilde{z})\right] \rho(\tilde{z}) d \tilde{z}+b_{1}\left(u \pm z_1\right)+b_{1}\left(u \pm z_2\right)+2 b_{1}(u)\no \\[6pt]
&&\qquad +b_{|2p|}(u)+2 b_{|2p+2|}(u)+b_{|2p+4|}(u)+b_{|2z_x-2|}(u)+2 b_{|2z_x|}(u)+b_{|2z_x+2|}(u) \no\\[6pt]
&&\quad = 2 N \int_{-\infty}^{\infty}\left[b_2(u-\theta)+b_4(u-\theta)\right] \sigma(\theta) d \theta \no \\[6pt]
 &&\qquad+b_{1}(u)+b_{3}(u)+b_{|2p|}(u)+b_{|2p+2|}(u)+b_{|2q|}(u)+b_{|2q+2|}(u),
\eea
where $\displaystyle b_n(u)=\frac{1}{2\pi}\frac{2u}{u^2+n^2/4}$. In the thermodynamic limit, the two pairs of real roots $\pm z_1, \pm z_2$ would tend to infinity and thus $b_1(u\pm z_1)+b_1(u\pm z_2)=0$. Eq.(\ref{intlam}) is a convolution equation and can be solved by the Fourier transformation
\bea
&&\hspace{-1.2truecm}\tilde{\rho}(w)\hspace{-0.05truecm}=\hspace{-0.05truecm}\tilde{\rho}_\delta(w)\hspace{-0.05truecm}+\hspace{-0.05truecm}\tilde{\rho}_{\rm {bstring }}(w),\\[6pt]
&&\hspace{-1.2truecm}\tilde{\rho}_\delta\hspace{-0.05truecm}=\hspace{-0.05truecm}\frac{2 N \tilde{\sigma}\left[\tilde{b}_2(w)\hspace{-0.05truecm}+\hspace{-0.05truecm}\tilde{b}_4(w)\right]\hspace{-0.05truecm}-\hspace{-0.05truecm}\tilde{b}_{1}(w)\hspace{-0.05truecm}+\hspace{-0.05truecm}\tilde{b}_{3}(w)\hspace{-0.05truecm}+\hspace{-0.05truecm}\tilde{b}_{|2p|}(w)\hspace{-0.05truecm}+\hspace{-0.05truecm}\tilde{b}_{|2p+2|}(w)\hspace{-0.05truecm}+\hspace{-0.05truecm}\tilde{b}_{|2q|}(w)\hspace{-0.05truecm}+\hspace{-0.05truecm}\tilde{b}_{|2q+2|}(w)}{N\left[\tilde{b}_{1}(w)\hspace{-0.05truecm}+\hspace{-0.05truecm}2 \tilde{b}_{3}(w)\hspace{-0.05truecm}+\hspace{-0.05truecm}\tilde{b}_{5}(w)\right]}, \\[6pt]
&&\hspace{-1.2truecm}\tilde{\rho}_{\rm {bstring }}(w)\hspace{-0.05truecm}=\hspace{-0.05truecm}-\hspace{-0.05truecm}\frac{\tilde{b}_{|2p|}(w)\hspace{-0.05truecm}+\hspace{-0.05truecm}2 \tilde{b}_{|2p+2|}(w)\hspace{-0.05truecm}+\hspace{-0.05truecm}\tilde{b}_{|2p+4|}(w)\hspace{-0.05truecm}+\hspace{-0.05truecm}\tilde{b}_{|2z_x-2|}(w)\hspace{-0.05truecm}+\hspace{-0.05truecm}2 \tilde{b}_{|2z_x|}(w)\hspace{-0.05truecm}+\hspace{-0.05truecm}\tilde{b}_{|2z_x+2|}(w)}{N\left[\tilde{b}_{1}(w)\hspace{-0.05truecm}+\hspace{-0.05truecm}2 \tilde{b}_{3}(w)\hspace{-0.05truecm}+\hspace{-0.05truecm}\tilde{b}_{5}(w)\right]},
\eea
where $\tilde{b}_n(w)=sign(k)ie^{-|nw|}$ and $\tilde{\rho}_{\rm {bstring }}(w)$ are the contributions of the boundary strings and extra roots to the density of the bulk strings $\tilde{\rho}_{\delta}(w)$. From now on, we use $\sigma(\theta)=\delta(\theta)$. The ground state energy of the Hamiltonian (\ref{H1}) in regime $B$ can thus be expressed as
\bea
E_{g} =&&\hspace{-0.4truecm}\frac{ N}{2} \int_{-\infty}^{\infty} d w \tilde{\rho}(w)\left[\tilde{a}_{5}(w)-\tilde{a}_{1}(w)\right]+4\no\\[6pt]
&&\hspace{-0.4truecm}-\frac{1}{p(p+1)}-\frac{1}{(p+1)(p+2)}-\frac{1}{z_x\left(z_x-1\right)}-\frac{1}{z_x\left(z_x+1\right)} \no\\[6pt]=&&\hspace{-0.4truecm}2 \pi-\frac{4}{3}+\frac{1}{p+1}-\frac{1}{p}+\frac{1}{q+1}-\frac{1}{q}-N,
\eea
where $\tilde{a}_n(k)=e^{-|nk|}$ is the Fourier transformation of $a_n(u)=\frac{1}{2\pi}\frac{n}{u^2+n^2/4}$.

Using the similar idea and after tedious calculations, we obtain the density $\tilde{\rho}_{\rm {bstring }}(w)$ in all the regimes of boundary parameters
\bea
\tilde{\rho}_{{\rm bstring }}=
\begin{cases}
0,\quad {\rm  in\hspace{4pt} regime }\hspace{4pt} A, \\
B_{p}+B_{p+1}+B_{z_x-1}+B_{z_x},\quad {\rm in\hspace{4pt} regime}\hspace{4pt} B, \\
B_{q}+B_{q+1}+B_{z_x-1}+B_{z_x},\quad {\rm  in\hspace{4pt} regime }\hspace{4pt} C, \\
B_{q}+B_{q+1}+B_{p}+B_{p+1},\quad {\rm  in\hspace{4pt} regime }\hspace{4pt} D, \\
B_{|q|},\quad {\rm  in\hspace{4pt} regime }\hspace{4pt} E, \\
B_{p}+B_{p+1}+B_{z_x-1}+B_{z_x}+B_{|q|},\quad {\rm  in\hspace{4pt} regime }\hspace{4pt} F, \\
B_{z_x-1}+B_{z_x}+B_{1-|q|}+2\cos (\lambda w)B_{1},\quad {\rm  in\hspace{4pt} regime }\hspace{4pt} G, \\
B_{p}+B_{p+1}+B_{1-|q|}+2\cos (\lambda w)B_{1},\quad {\rm  in\hspace{4pt} regime }\hspace{4pt} H, \\
B_{|q|-1}+B_{|q|},\quad {\rm in\hspace{4pt} regime }\hspace{4pt} I, \\
B_{p}+B_{p+1}+B_{z_x-1}+B_{z_x}+B_{|q|-1}+B_{|q|},\quad {\rm  in\hspace{4pt} regime }\hspace{4pt} J, \\
B_{z_x-1}+B_{z_x},\quad {\rm  in\hspace{4pt} regime }\hspace{4pt} K, \\
B_{p}+B_{p+1},\quad {\rm  in\hspace{4pt} regime }\hspace{4pt} L,
\end{cases}
\eea
where $B_a=-\frac{\tilde{b}_{2a}(w)+\tilde{b}_{2a+2}(w)}{N(\tilde{b}_{1}(w)+2\tilde{b}_{3}(w)+\tilde{b}_{5}(w))}$.
Substituting these solutions into the integration expression of the energy, we find that the ground state energies
can be expressed as
\bea
E_{g}  =
\begin{cases}
2 \pi-\frac{4}{3}+\frac{1}{p+1}-\frac{1}{p}+\frac{1}{q+1}-\frac{1}{q}-N,\quad p>0, q>0\hspace{4pt} \rm { or }\hspace{4pt} q<-1, \\[6pt]
2 \pi-\frac{4}{3}+\frac{1}{p+1}-\frac{1}{p}+\frac{1}{q+1}-\frac{1}{q}-N+2 \pi \csc (q \pi),\quad p>0, -1<p<0.
\end{cases}
\eea
Thus the surface energy $E_b$ of the system is
\bea
E_{b}  =
\begin{cases}
2 \pi-\frac{4}{3}+\frac{1}{p+1}-\frac{1}{p}+\frac{1}{q+1}-\frac{1}{q},\quad p>0, q>0\hspace{4pt} \rm { or }\hspace{4pt} q<-1, \\[6pt]
 2 \pi-\frac{4}{3}+\frac{1}{p+1}-\frac{1}{p}+\frac{1}{q+1}-\frac{1}{q}+2 \pi \csc (q \pi),\quad p>0, -1<p<0.
 \end{cases}
\eea
The surface energies with different boundary parameters $p$ and $q$ are shown in Fig.\ref{eb} below.

\begin{figure}[htp]
\centering
\includegraphics[width=12cm]{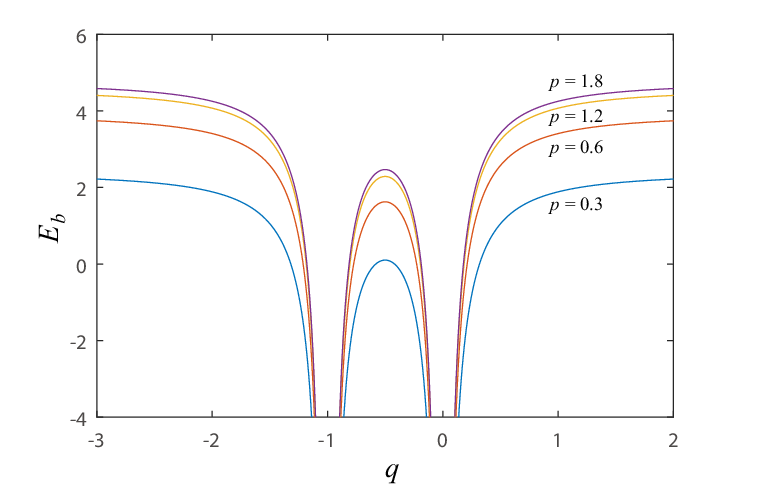}
  \caption{Surface energies versus the boundary parameters $p$ and $q$.}
\label{eb}
\end{figure}


\newpage
\section{Conclusions}
\label{Con}
In this paper, we have studied the thermodynamic limit and exact surface energy of the isotropic spin-1 Heisenberg chain with generic non-diagonal boundary.
The eigenvalues of the fused and fundamental transfer matrices are parameterized by their zero points, and the homogeneous zero points BAEs are given.
We have obtained the patterns and constraints of zero roots distributions in different regimes by solving the BAEs and analytical analysis.
Based on the patterns, the densities of zero roots and the exact surface energies in all regimes of the boundary parameters are derived.
The method and process presented in this paper can be generalized to the study of the spin-$s$ Heisenberg chain model. Results on this will be reported elsewhere.

\section*{Acknowledgments}

We thank Professor Yupeng Wang for valuable discussions. We acknowledge the financial support from National Key R$\&$D Program of China (Grant No.2021YFA1402104), Australian Research Council Discovery Project DP190101529 and Future Fellowship FT180100099,
China Postdoctoral Science Foundation Fellowship 2020M680724,
National Natural Science Foundation of China (Grant Nos. 12074410, 12047502, 12247179, 11934015 and 11975183), the Major Basic Research Program of Natural Science of Shaanxi Province
(Grant Nos. 2021JCW-19 and 2017ZDJC-32), and the Strategic Priority Research Program of the Chinese Academy of Sciences (Grant No. XDB33000000).


\appendix
\section{Fundamental spin-$(\frac{1}{2},1)$ transfer matrix $t^{\left(\frac{1}{2}, 1\right)}(u)$}
\setcounter{equation}{0}
\renewcommand{\theequation}{A.\arabic{equation}}
In this appendix we present some details for the fundamental spin-$(\frac{1}{2},1)$ transfer matrix $t^{\left(\frac{1}{2}, 1\right)}(u)$ (\ref{t1-21}). The fundamental $R$-matrix $R^{\left(\frac{1}{2}, 1\right)}(u)$ reads
\bea\label{r12}
R^{\left(\frac{1}{2}, 1\right)}(u)=u+\frac{\eta}{2}+\eta\vec \sigma_1\cdot \vec S_{2},
\eea
where $\vec{\sigma} (\sigma^{x},\sigma^{y},\sigma^{z})$ are the Pauli operators and $\vec S$ is given in (\ref{Sj}).
The $R$-matrix (\ref{r12}) enjoys QYBE and the following unitarity relation
\bea
R^{\left(\frac{1}{2}, 1\right)}_{12}(u)R^{\left(1,\frac{1}{2}\right)}_{21}(-u)= -(u+\frac{3}{2}\eta)(u-\frac{3}{2}\eta)\,\times{\rm id}.\label{Unitarity}
\eea
The fundamental reflection matrix $K^{-(\frac{1}{2})}(u)$ defined in the 2-dimensional (spin-1/2 representation) space is
\bea
K^{-(\frac{1}{2})}(u)&=&\left(
  \begin{array}{cccc}
    p_{-}+u &  \alpha_{-}u\\
    \alpha_{-}u  & p_{-}-u\\
  \end{array}
\right),
\eea
which satisfies the following RE
\bea
&&R_{12}^{\left(1,\frac{1}{2}\right)}(u-v)K_1^{-(1)}(u)R_{21}^{\left(\frac{1}{2},1\right)}(u+v)K_2^{-(\frac{1}{2})}(v)\no\\[6pt]
&&\hspace{1truecm}=K_2^{-(\frac{1}{2})}(v)R_{12}^{\left(1,\frac{1}{2}\right)}(u+v)K_1^{-(1)}(u)R_{21}^{\left(\frac{1}{2},1\right)}(u-v),
\eea
where $K^{-(1)}(u)$ is given in (\ref{K11f}). The fundamental dual reflection matrix $K^{+(\frac{1}{2})}(u)$ is
\bea
K^{+(\frac{1}{2})}(u)&=&K^{-(\frac{1}{2})}(-u-\eta)\Big|_{(p_{-},\alpha_{-})\rightarrow(p_{+},-\alpha_{+})}.
\eea

\end{document}